\documentclass[fleqn,usenatbib]{mnras}

\usepackage{newtxtext,newtxmath}
\usepackage[T1]{fontenc}

\DeclareRobustCommand{\VAN}[3]{#2}
\let\VANthebibliography\thebibliography
\def\thebibliography{\DeclareRobustCommand{\VAN}[3]{##3}\VANthebibliography}

\usepackage{graphicx}	% Including figure files
\usepackage{amsmath}	% Advanced maths commands
\usepackage{amssymb}	% Extra maths symbols

\title[The nature of the symbiotic candidate 2MASS J07363415+6538548]{The nature of the symbiotic candidate 2MASS J07363415+6538548 in~the~field of NGC~2403}

\author[J. Merc et al.]{
J. Merc,$^{1,2}$\thanks{E-mail: jaroslav.merc@student.upjs.sk}
R. G{\'a}lis,$^{2}$
J. K{\'a}ra,$^{1}$
M. Wolf,$^{1}$
and M. Vra\v{s}\v{t}\'{a}k$^{3}$
\\
$^{1}$Astronomical Institute, Faculty of Mathematics and Physics, Charles University, V Hole\v{s}ovi\v{c}k{\'a}ch 2, 180 00 Prague, Czech Republic\\
$^{2}$Institute of Physics, Faculty of Science, P. J. \v{S}af{\'a}rik University, Park Angelinum 9, 040 01 Ko\v{s}ice, Slovak Republic\\
$^{3}$ Private observatory Liptovsk\'{a} \v{S}tiavnica, K\v{l}u\v{c}iny 457/74, Slovak Republic\\
}

\date{Accepted 2020 September 25. Received 2020 September 11; in original form 2020 July 13.}

\pubyear{2020}

\begin{document}
\label{firstpage}
\pagerange{\pageref{firstpage}--\pageref{lastpage}}
\maketitle

\begin{abstract}
New Online Database of Symbiotic Variables includes several poorly characterised objects and candidate symbiotic stars, not only in the Milky Way but also in other galaxies. The goal of the research presented in this paper was to reveal the nature of 2MASS J07363415+6538548, the object discovered as an X-ray source in the field of NGC~2403, and tentatively classified as a symbiotic candidate or a cataclysmic variable. By analysis of available photometric data from ground-based surveys, together with a high precision photometry from \textit{TESS}, remarkable astrometric measurements of the \textit{Gaia} satellite and observations of other surveys spanning from X-rays to infrared, we have found that the object is not a symbiotic star nor a~cataclysmic variable but rather an active K-type dwarf. The star is located in the distance of 415\,pc, has an effective temperature of 4\,275\,K, luminosity of 0.14\,L$_{\sun}$, mass of $0.7$\,M$_{\sun}$, and radius of $0.7$\,R$_{\sun}$. It has a rotational period $\sim$\,3\,days and is a strong X-ray source with the X-ray luminosity of $\sim$\,$10^{30}\rm\,erg\,s^{-1}$. Gyrochronology and isochrone fitting confirmed that the star is young.
\end{abstract}


%%%%%%%%%%%%%%%%%%%%%%%%%%%%%%%%%%%%%%%%%%%%%%%%%%

%%%%%%%%%%%%%%%%% BODY OF PAPER %%%%%%%%%%%%%%%%%%

\section{Introduction}
Symbiotic binaries are strongly interacting systems, usually consisting of a cool giant of a spectral type M and a hot white dwarf \citep[see e.g. the reviews by][]{2012BaltA..21....5M, 2019arXiv190901389M}. These objects are unique astrophysical laboratories in the study of the mass transfer and accretion processes, stellar winds, jets, dust formation, or thermonuclear outbursts. They are also important for the models of the stellar evolution, e.g. they were proposed as one of the possible progenitors of supernovae Ia \citep[e.g.][and references therein]{2011A&A...530A..63P,2019MNRAS.485.5468I}.

Interesting features of symbiotic systems and their potential in studies of various astrophysical phenomena increased demand for a proper characterisation of the symbiotic population, resulting in a~systematic search for symbiotics in the Milky Way \citep[e.g.][]{2013MNRAS.432.3186M, 2014MNRAS.440.1410M} and in nearby galaxies \citep[e.g.][]{2008MNRAS.391L..84G,2012MNRAS.419..854G, 2015MNRAS.447..993G, 2014MNRAS.444..586M, 2017MNRAS.465.1699M,2018arXiv181106696I}.

As a response to the growing number of known symbiotic stars and promising candidates, we have created a new, modern and complex catalogue of these objects, the New Online Database of Symbiotic Variables\footnote{http://astronomy.science.upjs.sk/symbiotics/} \citep{2019RNAAS...3...28M}. Our aim is to have an updated database of confirmed symbiotic binaries with all available information on a particular object, together with the list of candidates which are suspected from the symbiotic nature based on various indicators such as a photometric behaviour, spectral appearance, or an \mbox{X-ray} activity. Several of known or suspected systems are, however, only poorly studied. One of the sub-objectives of the New Online Database of Symbiotic Variables is to clean the database from the misclassified objects. Such a task is important in order to understand the properties of the symbiotic population in the Milky Way and in other galaxies. 

One of such disputable symbiotic candidates is 2MASS J07363415+6538548 (hereafter 2M0736; other designations: UCAC4 779-021043; CXO J073634.1+653854; 1SXPS J073634.2+653853; 2SXPS 55518; 3XMM J073634.1+653855). This object was first detected by the \textit{Chandra X-Ray Observatory} thanks to its X-ray emission \citep{2003AJ....125.3025S}. Several authors attributed this source to the galaxy NGC~2403 \citep{2003AJ....125.3025S, 2010ApJ...725..842L,2010AJ....139.1066Y,2011ApJS..192...10L,2012MNRAS.419.2095M,2019MNRAS.483.5554E}, a spiral galaxy which belongs to the M81 group, located at the distance of 3.3\,Mpc \citep{2006Ap.....49....3K}. It was classified as a~\ion{H}{ii} region or a massive star cluster \citep{2003AJ....125.3025S}, a~potential supernova remnant \citep{2010ApJ...725..842L}, a high-mass X-ray binary \citep{2012MNRAS.419.2095M, 2017MNRAS.466.1019S}, and finally as a possible cataclysmic variable or a symbiotic binary \citep{2015AJ....150...94B}. 
We have analysed the available information on this object from the literature, data from the \textit{Gaia} DR2, photometry from various surveys, together with our own observations and claim that the object is a foreground, active red dwarf. The paper is organised as follows: in Section \ref{sec:observations}, we describe used observational data, and in Section \ref{sec:results}, we discuss the results concerning the distance to this object, its spectral appearance, photometric variability, and X-ray emission.

\section{Observational data}\label{sec:observations}

For the analysis of photometric variations of 2M0736, we have collected the photometric measurements from the All-Sky Automated Survey for Supernovae (ASAS-SN) \citep[\textit{V}~and \textit{g} filters; ][]{2014ApJ...788...48S, 2017PASP..129j4502K} obtained during 535 nights between JD\,2\,455\,950 and 2\,458\,995, and the Zwicky Transient Facility (ZTF) survey \citep[\textit{r}~and \textit{g} filters; ][]{2019PASP..131a8003M} obtained during 131 nights between JD 2\,458\,204 and 2\,458\,846. The angular resolution of ASAS-SN is $\sim$\,8"/pixel while that of ZTF is $\sim$\,1"/pixel. These data are supplemented by our observations in \textit{B, V, R} and \textit{I} filters, obtained on JD\,2\,458\,982 at the Liptovsk\'{a} \v{S}tiavnica Observatory, the Slovak Republic, using the Newtonian telescope 355/1600 equipped with CCD G2-1600, and on JD\,2\,459\,028 at Ond\v{r}ejov Observatory, the Czech Republic, using the Mayer 0.65-m (f/3.6) reflecting telescope equipped with the CCD G2-3200. Our data have similar angular resolution to that of the ZTF survey.

In addition, we have used the high precision data obtained by the \textit{Transiting Exoplanet Survey Satellite} \citep[\textit{TESS}; ][]{2015JATIS...1a4003R}. The angular resolution of the \textit{TESS} data is $\sim$\,21"/pixel and the photometric precision is about 1\% for the average magnitude of 2M0736. Our target was observed in 30 minute cadence mode in Sector 20 (JD\,2\,458\,842 -- 2\,458\,868). We have processed  Full Frame Images (FFIs) using the {\sc Lightkurve} Python package \citep{2018ascl.soft12013L}, allowing preparation of Target Pixel Files from FFIs, a background subtraction, aperture photometry, and removing of common systematic problems from the data.

For the analysis of spectral energy distribution (SED) of 2M0736, in addition to our data in $B$, $V$, $R$, and $I$ filters, we have collected the data from various catalogues and surveys: \textit{XMM-Newton} \citep[filters \textit{UVW1, UV};][]{2012MNRAS.426..903P}, Beijing-Arizona-Taiwan-Connecticut Sky Survey \citep[BATC; \textit{a, b, c, d, e, f, g, i, j, k, m, n, o, p};][]{2001ChJAA...1..372Z}, Panoramic Survey Telescope and Rapid Response System \citep[Pan-STARRS; \textit{g, r, i, z, y};][]{2016arXiv161205243F}, \textit{Gaia} \citep[\textit{$G_{BP}$, \textit{G}, \textit{$G_{RP}$}};][]{2018A&A...616A...1G}, Two Micron All-Sky Survey \citep[2MASS; \textit{J, H, K};][]{2006AJ....131.1163S}, \textit{Wide-field Infrared Survey Explorer} \citep[\textit{WISE}; \textit{W1, W2, W3};][]{2010AJ....140.1868W}, and \textit{Spitzer}/IRAC \citep[\textit{3.6$\mu$m, 4.5$\mu$m, 5.8$\mu$m, 8$\mu$m, 24$\mu$m;}][]{2015ApJS..219...42K}.

\begin{figure}
\centering
\includegraphics[width=\columnwidth]{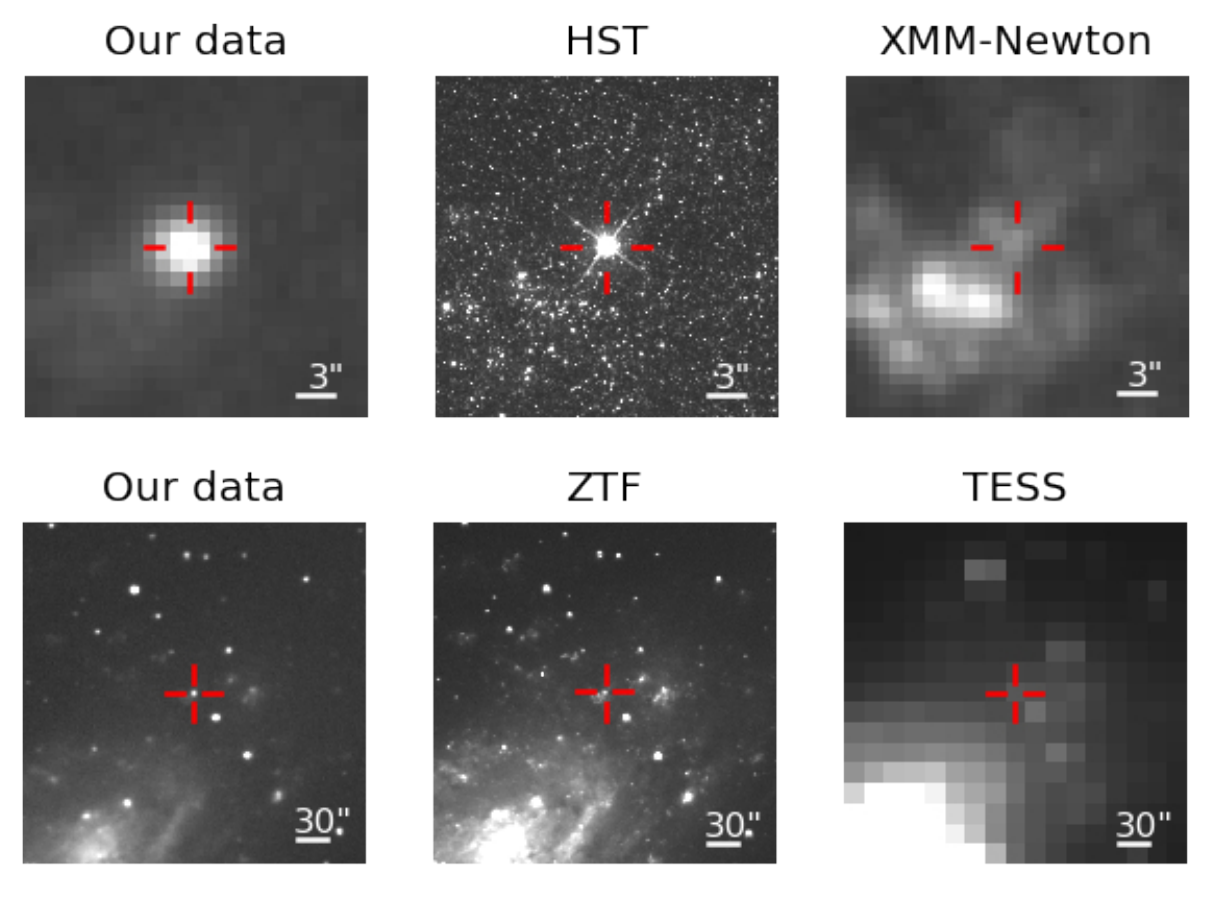}
\caption{Field of 2M0736. The upper panels show a comparison of 30" field from our observations ($V$ filter; Liptovsk\'{a} \v{S}tiavnica Observatory) with that of \textit{Hubble Space Telescope,} obtained on 19/8/2019 using the ACS Wide Field Channel in \textit{F814W} filter, and \textit{XMM-Newton} (filter \textit{UVW1}) acquired on 04/04/2014. The data were downloaded from the Hubble Legacy Archive \citep{2016AJ....151..134W} and the \textit{XMM-Newton} Science Archive, respectively. The lower panel shows comparison of the 360" field from our observations, the ZTF survey (filter \textit{g}) and \textit{TESS}.}
\label{fig:data}
\end{figure}

Fig. \ref{fig:data} shows a comparison of the field around 2M0736 based on our observations, compared to that of the \textit{Hubble Space Telescope}, the \textit{XMM-Newton}, the ZTF survey, and the \textit{TESS} mission. The influence of resolution on the data is well visible, especially in the case of \textit{TESS} observations, which have the lowest angular resolution. In this case, one should be cautious when interpreting the obtained results. We discuss this issue in Sec. \ref{sec:variability}. The image obtained by the \textit{Hubble Space Telescope} (Fig. \ref{fig:data}) revealed many faint stars located in the vicinity of our target, however, these are much fainter in the optical and infrared region than 2M0736. The opposite situation is true in the case of the UV data from \textit{XMM-Newton} as the nearby sources are more luminous in this spectral region than the studied star. Due to this fact, the fluxes in \textit{UVW1} and \textit{U} filters, presented in the XMM-Newton Serendipitous Ultraviolet Source Survey catalogue \citep{2012MNRAS.426..903P} are contaminated by the fluxes from nearby sources. We have remeasured fluxes in these two filters from the small, 4"~circular region around 2M0736 using the XMM-Newton Science Archive Interactive Analysis tool\footnote{http://nxsa.esac.esa.int/} and these values have been used for the analysis. It is possible that also the BATC \textit{a}~filter, centred on 3\,372\,\AA, is similarly affected by the nearby contaminants significant in this band-pass and therefore, we have excluded it from the analysis.

\section{Results and discussion}\label{sec:results}
In this section, the result concerning the distance, classification, optical variability, X-ray emission, and activity of 2M0736 are presented. The obtained basic physical parameters, including the age of the star under study, are also discussed. 

\subsection{Position, distance and reddening}\label{sec:distance}
The studied object, 2M0736, is located in the field of NGC 2403, in its northwest side, at $\alpha_{2000}$ = 07:36:34.14 and $\delta_{2000}$ = +65:38:54.86. As discussed in the introduction, several authors attributed the \mbox{X-ray} emission observed at this position to the source located in the galaxy itself. However, if this object is indeed a cataclysmic variable or a~symbiotic star, as proposed by \citet{2015AJ....150...94B}, it would be located in the Milky Way. \citeauthor{2015AJ....150...94B} have used the distance of 16\,kpc in the interpretation of their results, following \citet{2010PASP..122.1437P}, who obtained the spectral types and distances for stars in the Tycho2 catalogue by matching the Tycho2 $B_T$ and $V_T$, NOMAD $R_N$, and 2MASS $J$, $H$, and $K$ magnitudes with library spectra. 

\begin{table}
\caption{Basic properties of 2M0736. Data are based on our observations and the \textit{Gaia} DR2.}   \vspace{3mm}          % title of Table
\label{tab:gaia}      % is used to refer this table in the text
\centering            % used for centering table
\begin{tabular}{c c c c}  % centered columns (4 columns)
\hline  
Parameter               & Value           & Parameter  & Value            \\
\hline
$\alpha_{2000}$ [h:m:s] & 07:36:34.14     & $G$  [mag] & 15.43 $\pm$ 0.00 \\
$\delta_{2000}$ [d:m:s] & +65:38:54.86    & $BP$ [mag] & 16.15 $\pm$ 0.01 \\
l$_{2000}$ [deg]        & 150.515         & $RP$ [mag] & 14.56 $\pm$ 0.00 \\
b$_{2000}$ [deg]        & +29.158         & $B$  [mag] & 16.96 $\pm$ 0.08 \\
pm $\alpha$ [mas/yr]    &-2.38 $\pm$ 0.04 & $V$  [mag] & 15.74 $\pm$ 0.05 \\
pm $\delta$ [mas/yr]    & 5.98 $\pm$ 0.06 & $R$  [mag] & 15.07 $\pm$ 0.02 \\
parallax [mas]          & 2.36 $\pm$ 0.04 & $I$  [mag] & 14.33 $\pm$ 0.02 \\\hline
\end{tabular}
\end{table}

We have analysed the \textit{Gaia} DR2 data \citep{2018A&A...616A...1G} in order to confirm or reject the extragalactic nature of 2M0736. \textit{Gaia} data on our target are summarised in Table \ref{tab:gaia}. The measured values of proper motion (pm $\alpha = -2.38$\,mas/yr, pm $\delta = 5.98$\,mas/yr) and parallax ($\pi$ = 2.36) confirms, that 2M0736 is a~foreground star. Moreover, this measurements could be considered as reliable as the relative errors are low, < 2\% for all three values. 

It was found that the parallaxes from \textit{Gaia} DR2 are affected by the zero-point shifts \citep[$\sim$ 0.03 mas;][]{2018A&A...616A...2L,2018A&A...616A..17A}, which probably depends on the position in the sky and also the magnitude and colour of the object \citep{2018A&A...616A...2L}. However, this would not significantly affect the distance determination of our target as this becomes a significant issue only for the objects with values of measured parallaxes of the same order as the zero-point shift (i.e. distant objects).

When the relative error of the \textit{Gaia} DR2 parallax is reasonably low, inverting the parallax is applicable to obtain the accurate distance to the object \citep{2018A&A...616A...9L}. For 2M0736, this approach gives $d = 424$\,pc, which is very close to the distance obtained by \citet{2018AJ....156...58B} using the probabilistic, Bayesian approach based on the measured parallax and weak distance prior constructed according to the Milky Way model. For our target, they obtained the distance $d = 419$\,pc. Similarly, \citet{2019A&A...628A..94A} recovered the distance $d = 410$\,pc, employing, in addition to the geometrical assumptions, also the photometry from \textit{Gaia}, Pan-STARRS, 2MASS, and \textit{WISE}. Due to the distance determination errors of these estimates, which are around 8\,pc for all of the aforementioned methods, these results can be considered as identical. 
We have adopted the distance of $415\,^{+9}_{-5}$\,pc throughout this work. 

The \textit{Gaia} DR2 also provides the extinction estimate for the sub-sample of objects \citep{2018A&A...616A...1G}. However, this parameter is together with the effective temperature, radius, and luminosity of stars inferred only from the three broad-band photometric measurements (filters $G_{\rm BP}$, $G$,  $G_{\rm RP}$) and the value of the parallax (with the exception of $T\rm _{eff}$ which is based solely on the photometry). For this reason, one should be cautious when interpreting the obtained astrophysical parameters and the individual extinction estimates, which are rather poorly given for most of the stars in the \textit{Gaia} DR2 \citep{2018A&A...616A...8A}. For 2M0736, the value of $A\rm _G$ listed in the \textit{Gaia} DR2 is 0.465, corresponding to $E(B - V) = 0.19$\,mag.

In our analysis, we have used the extinction in the direction of 2M0736 obtained from the dust maps. We have employed the map by \citet{2011ApJ...737..103S} based on the measurements of the Sloan Digital Sky Survey, which gives the total galactic extinction in the direction to our target of $E(B - V) = 0.03$\,mag. We have also used the 3D map of interstellar dust reddening based on the photometry from Pan-STARRS and 2MASS published by \citet{2018MNRAS.478..651G}, which gives the extinction of $E(g - r) = 0.02$\,mag for the distance of 415\,pc. This corresponds to the value of $E(B - V) = 0.02$\,mag, under the assumption of $R\rm _V = 3.1$ and the extinction law of \citet{1989ApJ...345..245C}. For the purpose of the work, we have adopted the value of $E(B - V) = 0.03$\,mag. This relatively low value of extinction is further supported by the optical-infrared SED of the source.

\subsection{Classification of the source}
\begin{figure}
\centering
\includegraphics[width=\columnwidth]{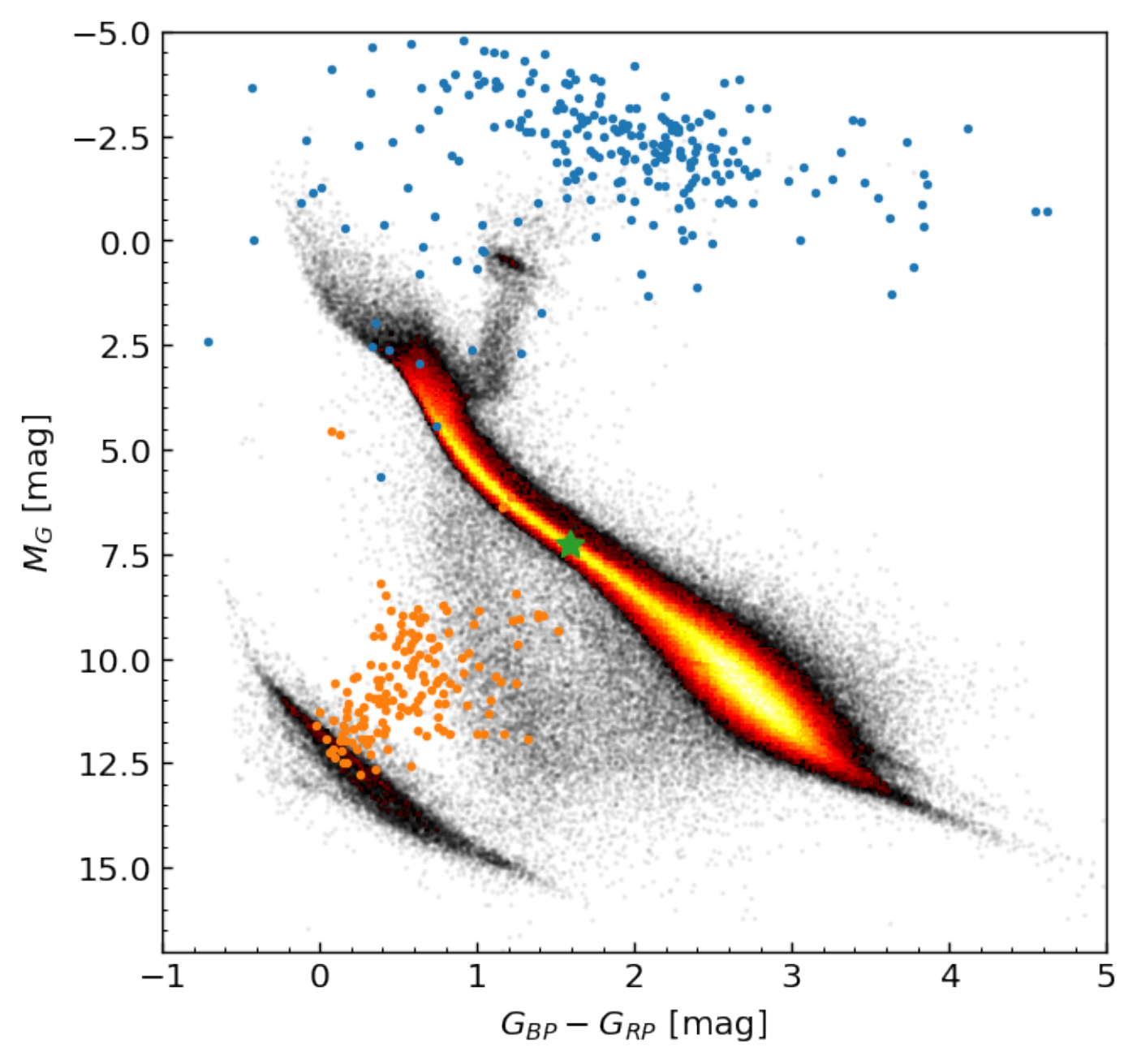}\vspace{-5mm}
\caption{Position of 2M0736 (green star symbol) in the \textit{Gaia} DR2 HR diagram of a sample of stars within 200\,pc with reliable astrometry \citep{2018A&A...616A..10G}. Known symbiotic variables from the New Online Database of Symbiotic Variables \citep{2019RNAAS...3...28M} and cataclysmic variables from the 150\,pc sample of \citet{2020MNRAS.494.3799P} together with the faint sample discovered by the SDSS survey \citep{2009MNRAS.397.2170G} are plotted with blue and orange symbols, respectively.}
\label{fig:hr}
\end{figure}

As the distance estimates presented in the previous section proved that the 2M0736 is located in the Milky Way rather than in NGC~2403, the further goal was to test the previously proposed classification of the object as a cataclysmic variable or a symbiotic star. The position of 2M0736 in the \textit{Gaia} HR diagram, together with the positions of the sample of cataclysmic variables and known symbiotic stars, is shown in Fig. \ref{fig:hr}. It is worth noting that these two groups are interacting binary systems. Therefore, the measured radiation in the \textit{Gaia} filters is a combination of the emission of both components of a binary and of sources connected with the interaction between the components, e.g. ionised nebula in the case of symbiotic stars or an accretion disc and its features (e.g. the hot spot) in case of cataclysmic variables.
% Two outlying cataclysmic variables are the nova-like variables.
% Tiež by som dodal, že "The outlying symbiotic stars will be the subject of further study."

One can see that 2M0736 is not located in the region occupied by any of these two groups of objects but resides on the main sequence (MS). Its position within MS suggests that it is a~red dwarf. From our observations, we have obtained the brightness $m \rm_V = 15.74$\,mag. From the apparent magnitude and the distance of 415\,pc, correcting for $E(B - V) = 0.03$\,mag, we have got absolute magnitude $M\rm _V = 7.56$\,mag, fully confirming the luminosity class V of our target.

We should note here that the classification of the star as a red dwarf is further supported by the observed optical variability (Sec. \ref{sec:variability}), and X-ray emission (Sec. \ref{sec:x-ray}). 

\subsection{Variability} \label{sec:variability}
\begin{figure}
\centering
\includegraphics[width=\columnwidth]{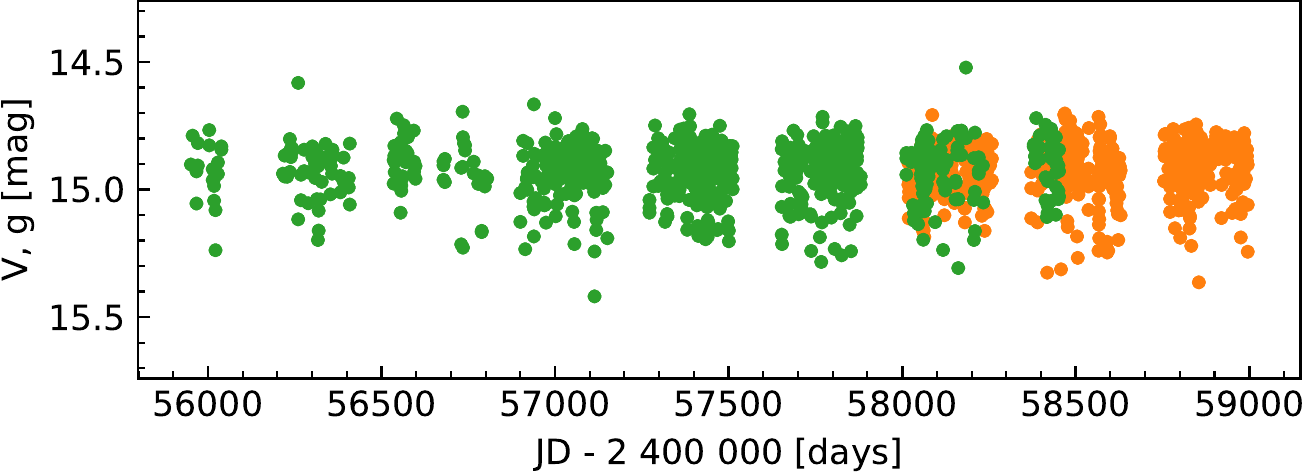}\vspace{-5mm}
\caption{Long-term light curves of 2M0736 from the ASAS-SN survey. Green and orange symbols represents the data in \textit{V} and \textit{g}, respectively. Data in the filter \textit{g} were shifted by - 0.22\,mag to match the average magnitude in the ASAS-SN data-set in \textit{V}.}
\label{fig:asasn_lc}
\end{figure}

To search for any variability of 2M0736, we have analysed its ASAS-SN, ZTF, and \textit{TESS} light curves. ASAS-SN observations span through the longest time period (3\,045\,days), so these are most useful for the analysis of long-term changes if any. We should note that the brightness of our object is close to the magnitude limit of the survey (typically the star was only <\,0.8\,mag brighter than the limiting magnitude of the particular observation). Moreover, the measurements of 2M0736 from ASAS-SN suffer from significant contamination probably due to lower angular resolution of the survey ($\sim$\,8"/pixel). The star had $m\rm _V \sim 15.7$\,mag during our observations, however, the average $V$ magnitude in the ASAS-SN survey is $\sim$ 14.9\,mag. In general, as seen in Fig. \ref{fig:asasn_lc}, no long-term trends are visible over the period of 2012 -- 2020, although the data are rather noisy.

\begin{figure}
\centering
\includegraphics[width=\columnwidth]{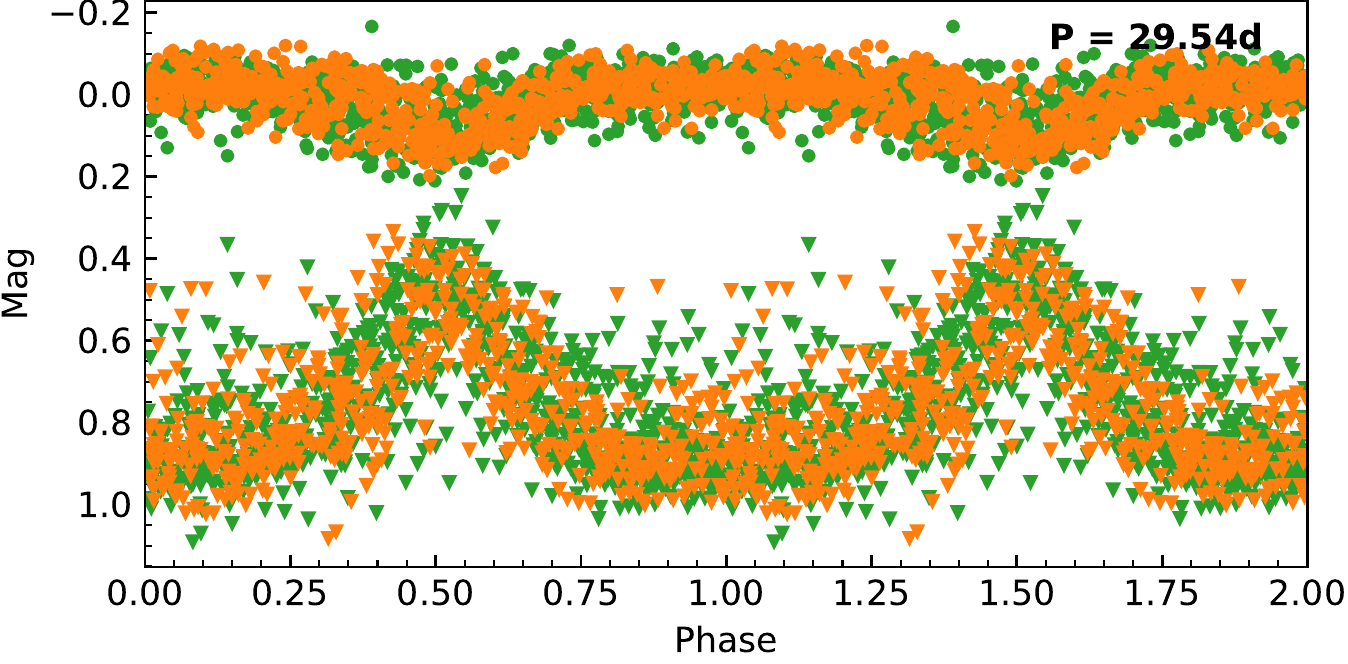}\vspace{-5mm}
\caption{Light curves of 2M0736 from the ASAS-SN survey phased with the period of 29.54\,days (dots symbols). Data in the filter \textit{g} were shifted by - 0.22\,mag to match the average magnitude in the ASAS-SN data-set in \textit{V}. Triangles represent the phased curves of limiting magnitudes for the same nights. Green and orange symbols represents the data in \textit{V} and \textit{g}, respectively. Note that the running average algorithm was applied to the light curves to reduce the noise. This period is not connected with the real variability of the object (see the text).}
\label{fig:asasn}
\end{figure}

Period analysis of the data was performed using the Lomb-Scargle and PDM methods in the software {\sc Peranso}\footnote{http://www.cbabelgium.com/peranso/}. The most significant period revealed in the data is one of 29.54 days. The ASAS-SN light curve of 2M0736 phased with this period is depicted in Fig. \ref{fig:asasn} (the upper curve). We have found that this period, closely resembling that of a synodic month (29.53 days), is not caused by the real variability of the object. Instead, it is probably related to the method of obtaining the data, specifically to the background subtraction, as the detected extremes in the light curve coincide with a~full moon. The same period was also obtained by the analysis of the limiting magnitudes given by the ASAS-SN for the particular observations (Fig. \ref{fig:asasn}, the lower curve). Moreover, we have not obtained such a period while using the ZTF and \textit{TESS} data (see below). Our period analysis of residues after removal of the 29.54 day period marginally detected also the presence of a shorter period 3.04 days. However, the amplitude of the variations with this period was of the same order as a typical error of the ASAS-SN measurements.

The most prominent period in the \textit{TESS} data is  one of 3.06\,days. The amplitude of the variations with this period ($\sim$\,0.1\,mag) was by an order greater than the typical measurement error from this survey ($\sim$\,0.01 mag for the average magnitude of 2M0736). On the other hand, the angular resolution of the \textit{TESS} data is rather low ($\sim$\,21"/px) and measured \textit{TESS} fluxes are surely contaminated by other nearby sources. Since the angular resolution of the ZTF survey is significantly higher ($\sim$\,1"/px), and the survey goes deeper (limiting magnitudes are 20.8\,mag and 20.6\,mag in \textit{g} and \textit{r} filter, respectively), we were able to identify these sources. The period analysis of their ZTF light curves does not show any variability with the period of 3.06\,days. Thus, it can be stated that the detected variations with the mentioned period of 3.06 days are related (with high probability) to the photometric variability of the studied star.

\begin{figure}
\centering
\includegraphics[width=\columnwidth]{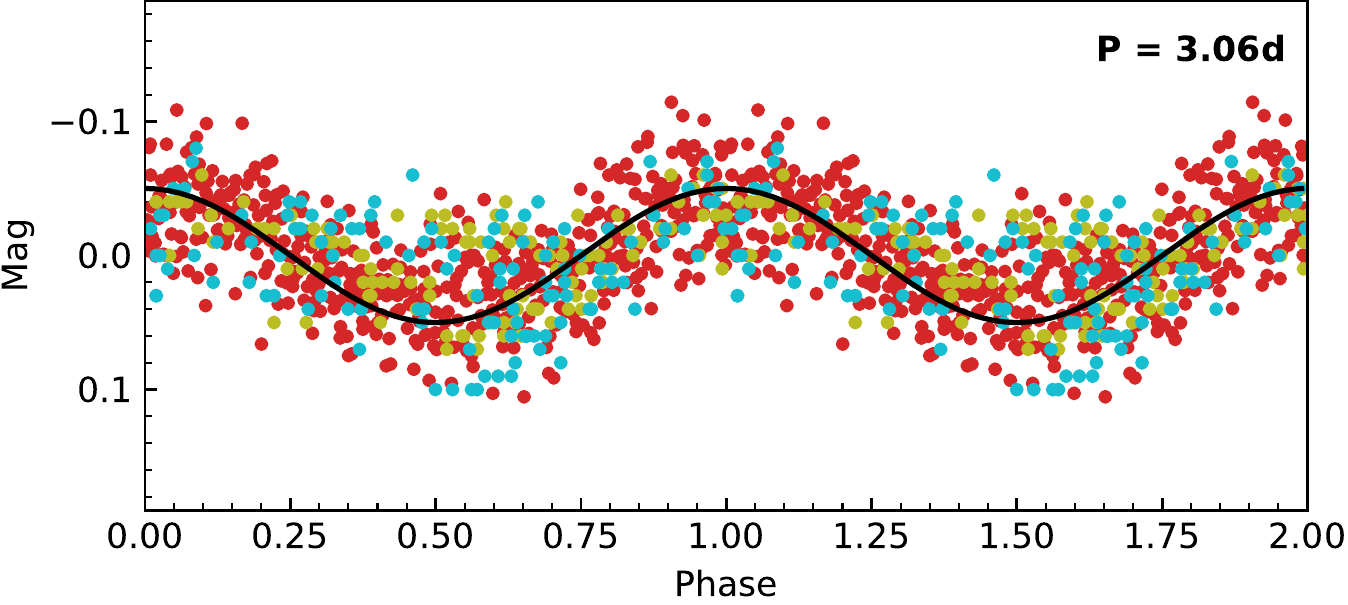}\vspace{-5mm}
\caption{Light curves phased with the period of 3.06\,days. Data obtained from \textit{TESS} are shown in red, from ZTF in the \textit{r} and \textit{g} filters by yellow and blue symbols, respectively. Daily averages were applied to the ZTF data for the dates with more than one observation for clarity. Black line represents a~sinusoidal variation with the given period.}
\label{fig:short}
\end{figure}

This result was also confirmed by our period analysis of the 2M0736 light curve constructed on the basis of the ZTF observations: the most prominent period was one of 3.06\,days. In addition to the same value of the period, the photometric variations in the \textit{TESS} and ZTF data had the same amplitude and phase, which made it possible to construct a joint phase diagram of the light curve of 2M0736 depicted in Fig. \ref{fig:short}. The simultaneous presence of photometric variations with the same period in all data sets confirms that they are indeed associated with the analysed object.

Such sinusoidal variability is common for red dwarfs. It is ascribed to their rotation and is due to the presence of starspot(s) on their surfaces as a result of their stellar activity. The regions covered by the starspots are cooler and consequently emit less flux, which results in the non-uniform surface brightness. As such star rotates, the minima in the light curve are observed when the starspot(s) face the observer. The rotation periods could be on the timescales of days to hundreds of days \citep[e.g.][]{2011ApJ...727...56I,2011ASPC..451..285E,2015ApJ...812....3W} and it was found that the younger dwarfs rotate faster \citep{1972ApJ...171..565S,2016Natur.529..181V,2017SoPh..292..126M}. The detected period value would classify 2M0736 among young red dwarfs, with the age of $\sim$\,0.2\,Gyr according to the calibration of \citet{2011ASPC..451..285E}. A~simple gyrochronology model implemented in the {\sc stardate} Python package \citep{2019AJ....158..173A} gives (from the \textit{Gaia} DR2 colours) the age of $\sim$\,0.08\,Gyr for this period. However, the implemented relations are probably not applicable to stars younger than a few hundred Myr. We have therefore employed a full MCMC approach together with isochrone fitting (see Sec. \ref{sec:spectral}).

\subsection{X-ray emission and activity} \label{sec:x-ray}
The studied object, 2M0736, has been reported to be a strong X-ray source by several authors (see the Introduction). It was studied in more detail by \citet{2015AJ....150...94B} using a single star model and the adopted distance of 16 kpc. For this distance, they obtained the X-ray luminosity of the source of 1.5\,$\times\,10^{33}\rm\,erg\,s^{-1}$ in 0.35 -- 8\,keV band. They concluded that the X-ray luminosity of the source is \mbox{2 -- 3} orders of magnitude higher in comparison with K-type giants from the catalogue of \citet{1998A&AS..127..251H}. At the same time, they argued that it is not consistent with the X-ray emission of main-sequence K stars based on the results of \citet{2009ApJS..181..444A}. For this reason, they preferred the binary model of a cataclysmic variable or a symbiotic system. 

We have re-calculated the X-ray luminosity of 2M0736 for the distance of 415 pc, which we have obtained in this study (see Sec. \ref{sec:distance}). We have inferred $L\rm _X$ = 1.0\,$\times\,10^{30}\rm\,erg\,s^{-1}$ for the unabsorbed flux $f\rm _X = 50.5\,\times\,10^{-15}\rm\,erg\,s^{-1}\,cm^{-2}$ presented in the work of \citet{2015AJ....150...94B}. In addition, we have re-analysed also the fluxes from the catalogues by \textit{Chandra} \citep{2010ApJS..189...37E}, \textit{XMM-Newton} \citep{2019yCat.9055....0R}, and \textit{Swift} \citep{2013yCat.9043....0E}. The unabsorbed fluxes in 0.35 -- 8\,keV band were estimated from the observed fluxes using the {\sc WebPIMMS} online tool\footnote{https://heasarc.gsfc.nasa.gov/cgi-bin/Tools/w3pimms/w3pimms.pl}, assuming $\Gamma = 1.9$. The Galactic column density for the position of our target, $N_{\rm H} = 3.96\,\times\,10^{20}\,\rm cm^{-2}$, was obtained from the map of \citet{2016A&A...594A.116H} using {\sc nH} web tool\footnote{https://heasarc.gsfc.nasa.gov/cgi-bin/Tools/w3nh/w3nh.pl}. From the unabsorbed fluxes, the X-ray luminosities of 8.9\,$\times\,10^{29}\rm\,erg\,s^{-1}$, 6.5\,$\times\,10^{29}\rm\,erg\,s^{-1}$, and 7.4\,$\times\,10^{29}\rm\,erg\,s^{-1}$ were obtained for the \textit{Chandra}, \textit{XMM-Newton}, and \textit{Swift} observations, respectively.

Using the \textit{V} magnitude of 2M0736 from our observations ($m_{\rm V}$ = 15.74\,mag), we have obtained the X-ray-to-optical flux ratio $\log{f_{\rm X}/f_{\rm V}}$ = -2.36, -2.49, and -2.43 for the \textit{Chandra}, \textit{XMM-Newton}, and \textit{Swift} observations, respectively. Calculated X-ray luminosities and the X-ray-to-optical flux ratios are consistent with an active, young K-type dwarf \citep[e.g.][]{2009ApJS..181..444A,2011A&A...531A...7G}. Moreover, such X-ray luminosity is expected to be observed for the dwarf rotating with the period of $\sim$\,3\,days (\citealp[see e.g. Fig. 3 and Fig. 5 in][]{2003A&A...397..147P}, or \citealp[Fig. 2 in][]{2020A&A...638A..20M}). 

The activity of such stars (e.g. photospheric starspots, chromospheric H$\alpha$ emission, and coronal X-rays) is probably caused by amplified local magnetic fields and depends on the rotation of the stars and the convective turbulence in their outer envelopes. Magnetic field generation in such stars occurs due to convective cells affected by the rotation \citep[e.g.][and references therein]{2020NatAs.tmp...46L}. This could be expressed in terms of the Rossby number, $R_{\rm o}$ = $P_{\rm rot}$/$\tau_{\rm c}$, the ratio of rotational period to convective turn-over time. The convective turn-over time cannot be directly deduced from the observations and needs to be estimated from the models of the stellar structure or using the empirical relations. Employing the empirical relation of convective turn-over time to $B - V$ colour by \citet{1984ApJ...279..763N}, we obtained $\tau_c$ = 24.5\,days. Using the relation of the stellar mass (derived in Sec. \ref{sec:spectral}) and $\tau_{\rm c}$ from \citet{2018MNRAS.479.2351W}, we got $\tau_{\rm c}$ = 28.4\,days. These values correspond to $R_{\rm o} = 0.11 - 0.12$. Derived values of the Rossby number puts the star close to the border between saturated and non-saturated regime in the activity-rotation relation \citep{2011ApJ...743...48W}.

The convective motion itself would have only an insignificant impact on the light curve of the studied object, as the predicted period range of oscillations caused by convection in dwarf stars is on the order of few tens of minutes and the amplitudes are probably too small (below few $\mu$mag) to be detected by present ground-based and space instruments \citep[see e.g.][]{2014MNRAS.438.2371R,2016MNRAS.457.1851R,2017MNRAS.469.4268B}.

\subsection{Spectral appearance and parameters of the star}\label{sec:spectral}

\begin{figure}
\centering
\includegraphics[width=\columnwidth]{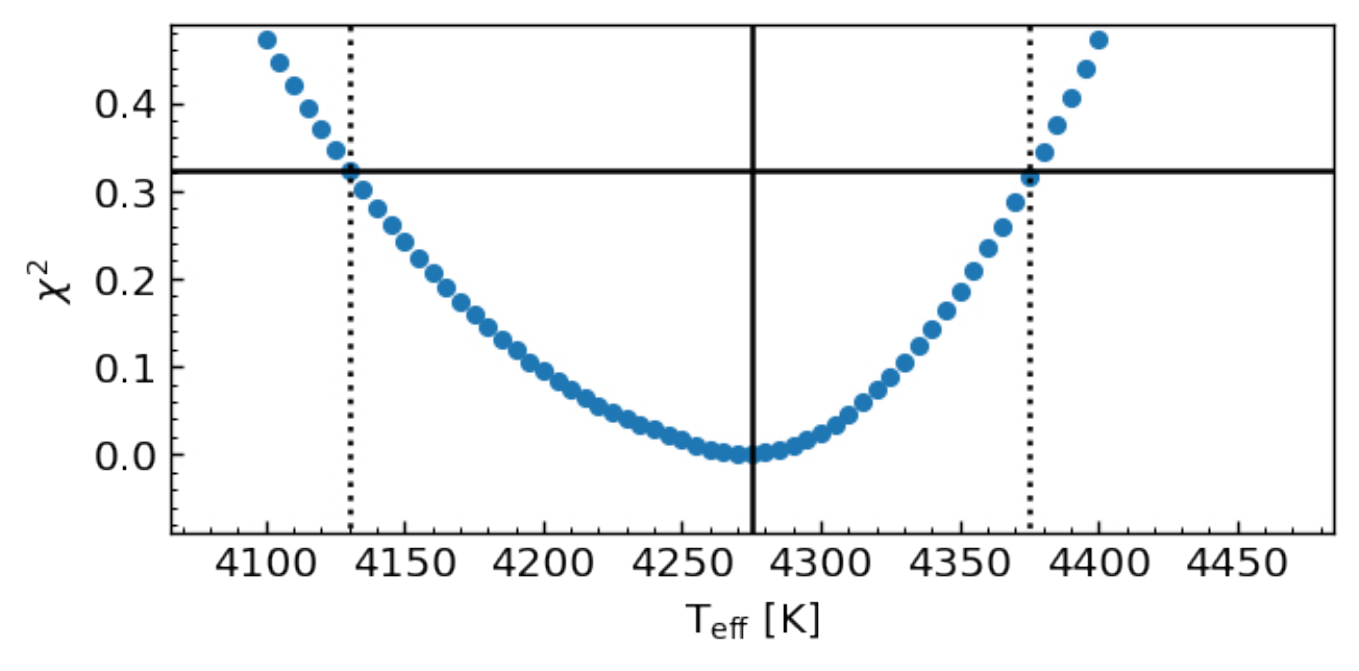}\vspace{-5mm}
\caption{Dependence of $\chi^2$ on $T\rm_{eff}$. Only a part of the full interval of tested effective temperatures is shown. The solid vertical line corresponds to the dependence minimum for $T_{\rm eff} = 4\,275$\,K. The solid horizontal line represents the sum of squared flux errors in all filters used for the calculation of the effective temperature uncertainty interval. The borders of this interval are depicted by the dotted vertical lines.}
\label{fig:chi}
\end{figure}

We have collected flux measurements of 2M0736 obtained in the UV, optical and IR spectral regions to supplement our observations in the optical filters \textit{B}, \textit{V}, \textit{R}, and \textit{I}. These data were used to estimate the astrophysical parameters of the star. To put constrains on the stellar parameters, especially $T\rm_{eff}$, and $\log{g}$, we have used the calibrations presented in \citet{2013JKAS...46..103S} and \citet{2018MNRAS.479.5491E}. Employing the absolute magnitude $M_{\rm V} = 7.56$\,mag, and colour of the star $(B-V)_0 = 1.19$\,mag, we have obtained the effective temperatures 4\,270\,K and 4\,260\,K, respectively. The value of $\log{g}$ derived from $M_{\rm V}$, and $(B - V)_0$ is $\sim$ 4.6 for both calibrations. 

Based on these results, we have prepared a grid of the \mbox{ATLAS9} model spectra by \citet{2003IAUS..210P.A20C} for effective temperatures $T\rm_{eff} = 3\,500 - 5\,000$\,K, with a step of 5\,K. We have used the fixed value of $\log{g} = 4.6$, as its influence on the shape of the continuum spectrum is negligible. With the availability of only spectro-photometry, we have adopted a solar metallicity, which should be a~sufficiently good approximation for a nearby, young dwarf.

\begin{figure}
\centering
\includegraphics[width=\columnwidth]{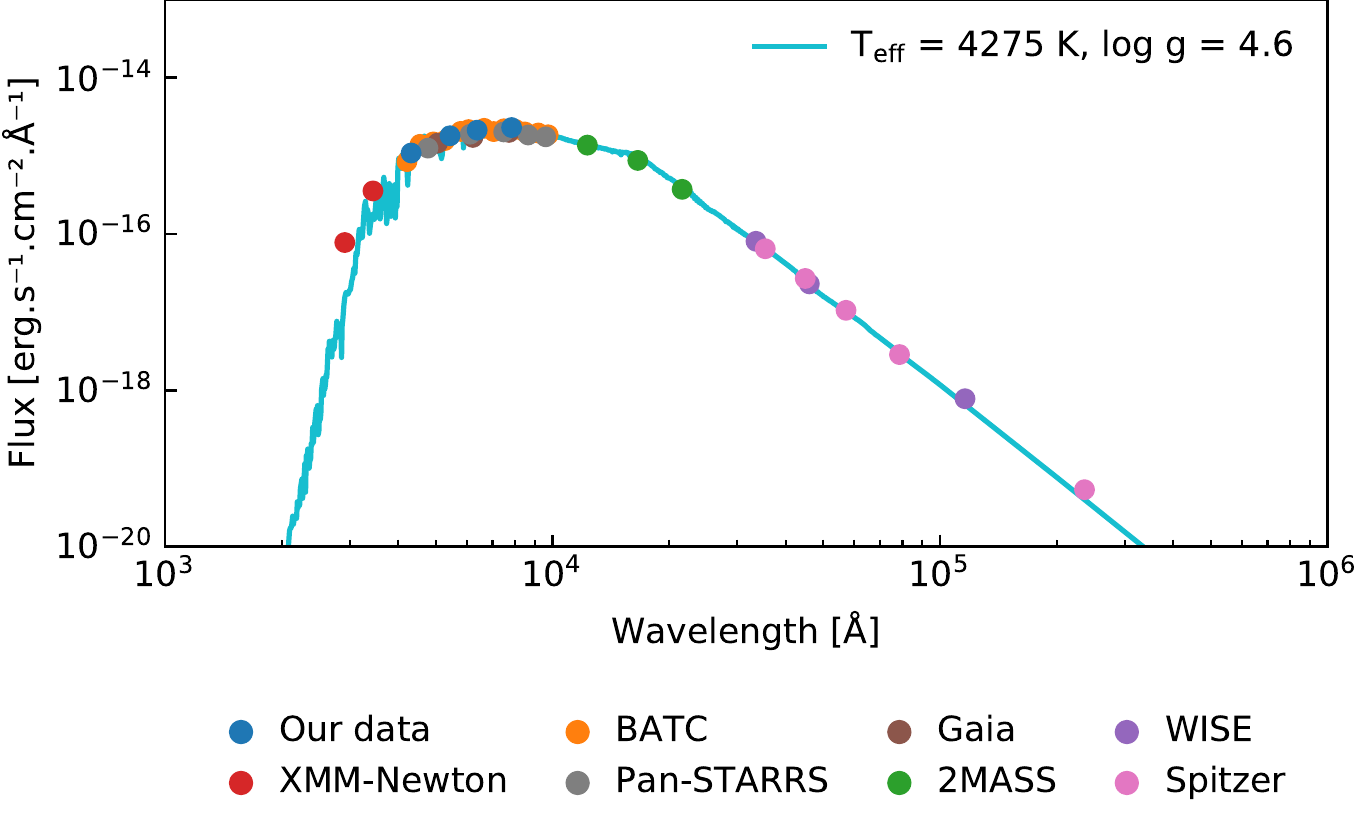}\vspace{-5mm}
\caption{Optical-infrared SED of 2M0736 together with the reddened \mbox{ATLAS9} stellar atmosphere model for $T_{\rm eff} = 4\,275$\,K and $\log{g}$ = 4.6 \citep{2003IAUS..210P.A20C}. The plotted spectro-photometry is constructed based on our data in \textit{B, V, R,}, and \textit{I} filters and supplemented with the data obtained by \textit{XMM-Newton}, the BATC survey, Pan-STARRS, \textit{Gaia}, \textit{2MASS}, \textit{WISE}, and \textit{Spitzer}/IRAC.}
\label{fig:sed}
\end{figure}

Each model spectrum was reddened by the value of the excess $E(B - V) = 0.03$\,mag, assuming $R_{\rm V}$ = 3.1 and the reddening law of \citet{1989ApJ...345..245C}, and normalised to the measured flux of 2M0736 in the \textit{J} filter. We then computed synthetic photometric fluxes for every spectrum in \textit{B, V, R, I, J, H, K, W1, W2}, and \textit{W3} filters and compared these values with observed ones in these filters. The $\chi^2$ value obtained this way was used to characterise the quality of the fit. The dependence of $\chi^2$ on the $T\rm_{eff}$ is plotted in Fig. \ref{fig:chi}. The minimum of the dependence was reached for the value of $T_{\rm eff} = 4\,275$\,K, very close to the values obtained from the calibrations. This confirms that such relations work very well in this part of the MS. The derived value of the effective temperature corresponds to the spectral type $\sim$\,K6 \citep{2018MNRAS.479.5491E,2020arXiv200407627M}. The optical-infrared SED of 2M0736, together with the best fitting ATLAS9 model spectrum, is shown in Fig. \ref{fig:sed}.

To obtain an estimate of the effective temperature uncertainty interval, we have used observational errors of the fluxes of 2M0736 in the individual filters. The sum of the squared flux errors limits plausible effective temperatures and provides the interval \mbox{$T_{\rm eff} = 4\,130 - 4\,375 $\,K.} This range of temperatures was used to obtain the errors on other astrophysical parameters of the star. For $\log{g}$, the effective temperature interval corresponds to values $4.58 - 4.64$. We have therefore adopted the error of 0.04.

Using the aforementioned calibrations of \citet{2018MNRAS.479.5491E}, which are based on the mass-luminosity, mass-temperature and mass-radius relations, we obtained a mass of the 2M0736 star, $M = 0.71\pm{0.04}$\,M$_{\sun}$, and its radius $R = 0.70\pm{0.05}$\,R$_{\sun}$. Consequently, from $L = 4\pi R^2\sigma T^{4}_{\rm eff}$, we obtained a luminosity of $0.15\pm{0.03}$\,L$_{\sun}$. Very similar results could be achieved by using the calibration of \citet{2013ApJS..208....9P} for the temperature of 4\,275\,K: $M = 0.65$\,M$_{\sun}$, $R = 0.67$\,R$_{\sun}$, and $L = 0.13$\,L$_{\sun}$. We should note that also the colours and the absolute magnitudes for the obtained astrophysical parameters are in very good agreement with the observed values.

\begin{figure}
\centering
\includegraphics[width=\columnwidth]{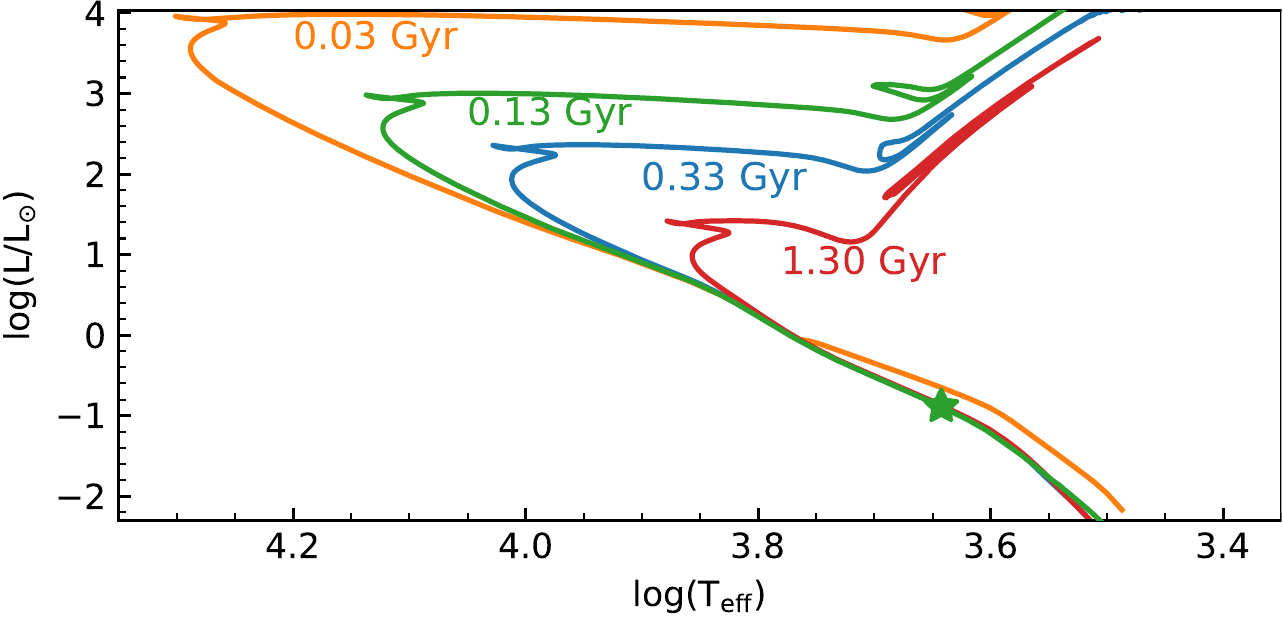}\vspace{-5mm}
\caption{MESA isochrones \citep{2016ApJ...823..102C} for ages 0.03 (orange), 0.13 (green), 0.33 (blue), and 1.30 Gyr (red). The position of 2M0736 is shown by green star symbol. The value of 0.13 Gyr for the star was obtained combining isochrone fitting and gyrochronology. Values of age orders of magnitude higher are not preferred due to short rotational period (see the text).}
\label{fig:isochrone}
\end{figure}

In order to obtain the age estimate of the star, the derived parameters $T_{\rm eff}$, $\log{g}$, and $L$, together with the \textit{Gaia} DR2 parallax, the extinction $A_{\rm V}$, \textit{B} and \textit{V} filters measurements, and the calculated rotational period, were used as the input parameters for combined isochrone MCMC fitting and gyrochronology, using the {\sc stardate} Python package \citep{2019AJ....158..173A}. We performed $5 \times 10^5$ iterations and obtained the age of $0.13^{+0.08}_{-0.10}$\,Gyr. Other parameters of 2M0736 acquired from the isochrone fitting are as follows: $T_{\rm eff} = 4\,391\,$K, $\log{g} = 4.67$, [Fe/H] = -0.04, $M = 0.68$\,M$_{\sun}$, $R = 0.63$\,R$_{\sun}$, and the distance of 424\,pc. We should note, that standalone isochrone fitting (without the gyrochronology) would yield enormous errors in this case, as the isochrones are very close to each other in this part of the HR diagram (see Fig. \ref{fig:isochrone}). The MESA (Modules for Experiments in Stellar Astrophysics) isochrones shown in the figure for ages 0.03, 0.13, 0.33, and 1.3 Gyr are calculated for solar metallicity using {\sc MIST} Web Interpolator\footnote{http://waps.cfa.harvard.edu/MIST/interp\_isos.html} \citep{2016ApJ...823..102C}.

As we have mentioned in Sec. \ref{sec:distance}, the \textit{Gaia} DR2 catalogue \citep{2018A&A...616A...1G} provides an estimate of astrophysical parameters, too. However, these are based only on the broad-band photometry, and the parallax, with the exception of the effective temperature, which is solely based on the photometric measurements. Still, the given temperature, 4\,367\,K (in the uncertainty interval 4\,197 - 4\,795\,K) is similar to ours, taking into account that extinction used in the \textit{Gaia} DR2 is slightly higher than the value we have used. Radius of the star in the catalogue is $\sim$\,0.61\,R$_{\sun}$ (0.51 - 0.66\,R$_{\sun}$) and the luminosity is $\sim$\,0.12\,L$_{\sun}$ (0.12 - 0.13\,L$_{\sun}$), consistent with our results. The values in brackets denote 16th and 84th percentile. Very similar results, $T_{\rm eff}$ = 4\,508\,K (4\,404 - 4607\,K), $\log{g}$ = 4.62 (4.61 - 4.64), [Fe/H] = -0.02 (-0.13 - 0.10), and $M = 0.70$\,$M_{\sun}$ (0.65 - 0.70\,$M_{\sun}$) were obtained by \citet{2019A&A...628A..94A}, employing the \textit{Gaia} DR2 data, together with the photometric measurements from Pan-STARRS, 2MASS, and \textit{WISE} (see also Sec. \ref{sec:distance}). Also in this case, the reason for the higher value of the temperature would be the higher adopted value of the extinction. Finally, \citet{2019AJ....158...93B} obtained the temperature of 4\,260$\pm$161\,K from the \textit{Gaia} DR2 data (position, parallax, proper motions, and magnitudes), employing the machine learning regression technique trained with data from LAMOST, SSPP, RAVE and APOGEE catalogues. For clarity, we have summarised the results obtained in this work and by other authors in Table \ref{tab:astropar}.

The advantage of the results of our analysis based on SED modelling and isochrone fitting with gyrochonology is that we obtained all the stellar parameters of the investigated object in contrast to the limited amount of parameters obtained from the catalogues. Our modelling is also based on observational data covering a wider spectral interval (towards both shorter and longer wavelengths) than in the case of these external sources. In contrast to these catalogues, an independently-defined extinction value for 2M0736 was also used in our analysis (see Sec. \ref{sec:distance}).

\begin{table*}
\caption{Comparison of the stellar parameters of 2M0736 obtained in this work by the SED fitting/calibrations (SED) and isochrone fitting with gyrochronology (ISO), together with the values presented in the Gaia DR2 catalogue \citep[DR2;][]{2018A&A...616A...1G}, and the ones obtained from the {\sc StarHorse} catalogue \citep[SHC;][]{2019A&A...628A..94A} and from the catalogue of \citet{2019AJ....158...93B} acquired using the machine learning regression (MLR) technique.}   

\vspace{3mm}          % title of Table
\label{tab:astropar}      % is used to refer this table in the text
\centering 
\begin{tabular}{lccccc}
\hline  
 & \multicolumn{2}{c}{This work} & DR2 & SHC & MLR \\
 & SED & ISO &  &  &  \\
 \hline  
\vspace{1mm}
$T_{\rm eff}$ [K] & $4\,275 \,^{+100} _{-145}$ & 4\,391\,$^{+33}_{-81}$ & $4\,367\,^{+428} _{-170}$ & 4\,508\,$^{+99}_{-103}$ & 4\,260$\pm$161 \\
\vspace{1mm} 
$\log{g}$ & $4.60 \, ^{+0.04} _{-0.02}$ & 4.67\,$^{+0.01}_{-0.07}$ & - & 4.62\,$^{+0.02}_{-0.01}$ & - \\
\vspace{1mm}
$L$ [L$_{\sun}$] & $0.14 \, ^{+0.03} _{-0.03}$ & 0.13\,$^{+0.03}_{-0.01}$ & $0.12\,^{+0.01} _{-0.01}$ & - & - \\
\vspace{1mm}
$M$ [M$_{\sun}$]& $0.68 \, ^{+0.04} _{-0.04}$ & 0.68\,$^{+0.02}_{-0.01}$ & - & 0.70\,$^{+0.01}_{-0.05}$ & - \\
\vspace{1mm}
$R$ [R$_{\sun}$]& $0.69 \, ^{+0.05} _{-0.05}$ & 0.63\,$^{+0.06}_{-0.01}$ & $0.61\,^{+0.05} _{-0.10}$ & - & -\\\hline
\end{tabular}
\end{table*}

\section{Conclusion}
In this work, we have analysed the available astrometric, photometric, and spectro-photometric data of 2MASS J07363415+6538548, an object located in the field of NGC~2403. It was previously suspected of being a symbiotic star and listed as a symbiotic candidate in the New Online Database of Symbiotic Variables. We have shown that the object is actually located in the Milky Way, in the distance of 415\,pc and that it is an active, young K-type dwarf rather than a cataclysmic variable or a symbiotic binary. This classification is confirmed by the position of the object in the HR diagram and by analysis of its spectral energy distribution. 

From the relations of astrophysical parameters on the absolute magnitude and colour of the object, combined with the SED fitting, gyrochronology, and isochrone fitting, we have obtained its effective temperature of 4\,275\,K, luminosity of 0.14\,L$_{\sun}$, mass of $0.7$\,M$_{\sun}$, radius of $0.7$\,R$_{\sun}$, and the age of 0.13 Gyr. The rotational period of $\sim$\,3\,days, obtained using the data from the \textit{TESS} satellite and the ZTF survey, and the X-ray luminosity of $\sim$\,$10^{30}\rm\,erg\,s^{-1}$, observed by \textit{Chandra}, \textit{XMM-Newton} and \textit{Swift}, are the expected values for a~young dwarf star. 

No variability of the object with a longer period has been detected. However, an artificial period of $\sim$\,29.5\,days is present in the ASAS-SN data. This period is probably connected with the background subtraction procedure, as it is equal to the length of the synodic month. This result could be important when analysing the variability of objects with brightness close to the survey limit, as it is in our case. 

Although the studied star has not been confirmed as a symbiotic star, this results is very important for the study of the whole symbiotic population based on our New Online Database of Symbiotic Variables. There surely are several other objects classified as symbiotic candidates, which are not symbiotic stars at all, and such objects might also be among those confirmed as symbiotic stars in the literature \citep[see e.g.][]{2020arXiv200410407A}. 

The presented research is one of the first steps to provide a clean sample of the symbiotic population to use it for comprehensive studies. As we have shown here, the position in the \textit{Gaia} DR2 HR diagram, analysis of the photometric variability, or unusual X-ray-to-optical flux ratios could be applied in addition to traditionally used near IR color-color diagrams and spectroscopic criteria \citep[e.g.][]{2000A&AS..146..407B,2017A&A...606A.110I} to identify disputable symbiotic candidates in the present sample. At the same time, these allow to select the most promising candidates for detailed spectroscopic follow-up. 

\section*{Acknowledgements}
We are thankful to an anonymous referee for the comments and suggestions greatly improving the manuscript. This research was supported by the \textit{Slovak Research and Development Agency} under contract No. APVV-15-0458, by the \textit{Charles University}, project GA UK No. 890120 and by the internal grant VVGS-PF-2019-1047 of the \textit{Faculty of Science, P. J. \v{S}af\'{a}rik University in Ko\v{s}ice}.

Based on observations obtained with the Samuel Oschin 48-inch Telescope at the Palomar Observatory as part of the Zwicky Transient Facility project. ZTF is supported by the National Science Foundation under Grant No. AST-1440341 and a collaboration including Caltech, IPAC, the Weizmann Institute for Science, the Oskar Klein Center at Stockholm University, the University of Maryland, the University of Washington, Deutsches Elektronen-Synchrotron and Humboldt University, Los Alamos National Laboratories, the TANGO Consortium of Taiwan, the University of Wisconsin at Milwaukee, and Lawrence Berkeley National Laboratories. Operations are conducted by COO, IPAC, and UW. This work has made use of data from the European Space Agency (ESA) mission {\it Gaia} (\url{https://www.cosmos.esa.int/gaia}), processed by the {\it Gaia} Data Processing and Analysis Consortium (DPAC, \url{https://www.cosmos.esa.int/web/gaia/dpac/consortium}). Funding for the DPAC has been provided by national institutions, in particular the institutions participating in the {\it Gaia} Multilateral Agreement; the data collected with the \textit{TESS} mission, obtained from the MAST data archive at the Space Telescope Science Institute (STScI). Funding for the \textit{TESS} mission is provided by the NASA Explorer Program. STScI is operated by the Association of Universities for Research in Astronomy, Inc., under NASA contract NAS 5–26555; the data products from the Two Micron All Sky Survey, which is a joint project of the University of Massachusetts and the Infrared Processing and Analysis Center/California Institute of Technology, funded by the National Aeronautics and Space Administration and the National Science Foundation and the data products from the Wide-field Infrared Survey Explorer, which is a joint project of the University of California, Los Angeles, and the Jet Propulsion Laboratory/California Institute of Technology, funded by the National Aeronautics and Space Administration. We acknowledge the use of the observations made with the NASA/ESA Hubble Space Telescope, and obtained from the Hubble Legacy Archive, which is a collaboration between the Space Telescope Science Institute (STScI/NASA), the Space Telescope European Coordinating Facility (ST-ECF/ESA) and the Canadian Astronomy Data Centre (CADC/NRC/CSA). 
We acknowledge the use of the SIMBAD and VIZIER databases, operated at CDS, Strasbourg, France. 

%%%%%%%%%%%%%%%%%%%%%%%%%%%%%%%%%%%%%%%%%%%%%%%%%%

\section*{Data availability}
The data underlying this article are available in the article.

\bibliographystyle{mnras}
\bibliography{mnras} % if your bibtex file is called example.bib

%%%%%%%%%%%%%%%%%%%%%%%%%%%%%%%%%%%%%%%%%%%%%%%%%%

%%%%%%%%%%%%%%%%% APPENDICES %%%%%%%%%%%%%%%%%%%%%

%\appendix

%%%%%%%%%%%%%%%%%%%%%%%%%%%%%%%%%%%%%%%%%%%%%%%%%%

% Don't change these lines
\bsp	% typesetting comment
\label{lastpage}
\end{document}